\documentstyle[epsf]{mn}

\def\Rg{r_{\rm g}}
\def\Rin{R_{\rm in}}

\def\LX{L_{\rm X}}
\def\lh{l_{\rm h}}
\def\ls{l_{\rm s}}

\def\TIC{T_{\rm IC}}
\def\Ts{T_{\rm s}}
\def\Ttrans{T_{\rm trans}}
\def\Teff{T_{\rm eff}}

\def\me{m_{\rm e}}

\def\nH{n_{\rm H}}
\def\ne{n_{\rm e}}

\def\MEdd{\dot M_{\rm Edd}}

\def\xte{{\it RXTE}\ }
\def\OmegaK{\Omega_{\rm K}}
\def\FX{F_{\rm X}}

\def\sigmaT{\sigma_{\rm T}}

\def\FD{F_{\rm D}}
\def\Fdown{F_{\rm refl}}
\def\Fsoft{F_{\rm soft}}
\def\Fhard{F_{\rm hard}}
\def\Fup{F_{\rm direct}}
\def\Htot{H_{\rm tot}}
\def\Hd{H_{\rm d}}

\def\Ftot{F_{\rm tot}}

\def\Pgas{P_{\rm gas}}
\def\Prad{P_{\rm rad}}
\def\Ptot{P_{\rm tot}}
\def\Prefl{p_{\rm rf}}
\def\Ptrans{p_{\rm tr}}
\def\MSun{{\rm M}_{\odot}}
\def\tdyn{t_{\rm dyn}}
\def\rmin{r_{\rm min}}
\def\rmax{r_{\rm max}}
\def\rcut{r_{\rm cut}}
\def\rms{r_{\rm ms}}
\def\lx{l_{\rm X}}
\def\fc{f_{\rm c}}
\def\Al{A_{\rm l}}
\def\tauh{\tau_{\rm hot}}
\def\sigvar{\sigma_{\rm var}}
\def\mcg{MCG--6-30-15\ }
\def\Ka{K$_{\alpha}$\ }
\def\d{{\rm d}}

\def\mus{\mu_{\rm s}}

\title[Variability of X--ray reprocessing in AGN]
{Testing a model of variability of X--ray reprocessing features in Active 
Galactic Nuclei}

\author[\.{Z}ycki, R\'{o}\.{z}a\'{n}ska]
{Piotr T. \.{Z}ycki\thanks{E-mail: ptz@camk.edu.pl, agata@camk.edu.pl},   
   Agata R\'{o}\.{z}a\'{n}ska
 \\
Nicolaus Copernicus Astronomical Center, Bartycka 18, 00-716 Warsaw,
Poland}

\date{13 February 2001}

\begin{document}

\maketitle

\begin{abstract}

A number of recent results from X--ray observations of Active Galactic Nuclei 
involving the Fe \Ka line 
(reduction of line variability compared to the X--ray continuum variability,
the X--ray ``Baldwin effect'') were attributed to a presence of a hot,
ionized skin of an accretion disc, suppressing emission of 
the line. The ionized skin appears as a result of the thermal
instability of X--ray irradiated plasma. 
We test this hypothesis by computing the Thomson
thickness of the hot skin on top of the $\alpha\Ptot$ Shakura--Sunyaev disc,
by simultaneously solving the vertical structure of both the hot skin and 
the disc. We then compute a number of relations between observable
quantities, e.g.\  the hard X--ray 
flux, amplitude of the observed reprocessed component, relativistic smearing
of the \Ka line, the r.m.s.\ variability of the hard X--rays. These
relations  can be compared to present and future observations. 
We point out that this mechanism  is unlikely to explain the behaviour of 
the X--ray source in MCG--6-30-15, where 
there is a number
of arguments against the existence of a thick hot skin, but it can work
for some other Seyfert 1 galaxies.

\end{abstract}


\section{Introduction}

Spectral features originating as a result of reprocessing of energetic X--ray
radiation by cool, optically thick plasma around accreting compact objects 
can be  extremely helpful in constraining the geometry of accretion flows.
Indeed, one of the strongest evidences for the existence of accretion discs
very close to central black holes in active galactic nuclei (AGN) came from
observations of broad profiles of the fluorescent iron \Ka line near 6.4 keV 
in many
Seyfert 1 galaxies, the {\it ASCA\/} result from \mcg (Tanaka et al.\ 1995)
being the best known case. 
The broad line profile is unlikely to be explained as 
a result of Comptonization in transmission through a plasma cloud (Fabian
et al.\ 1995), so broadening as a result of Doppler effects
and gravitational redshift seems to be the main 
mechanism (although Comptonization in reflection can contribute to the 
broadening; Karas et al.\ 2000).
Quantitative estimates of the distances of the cold plasma from central 
black holes based
on the line profiles indicate an intrinsic dispersion of parameters
among the sources 
(Nandra et al.\ 1997; Done, Madejski \& \.{Z}ycki 2000), but generally
yield the distance $< 100\Rg$ ($\Rg \equiv G M/c^2$).

Data analyzed in most cases are integrated over relatively long periods
of time ($\sim$days) in order to decrease statistical errors. They are 
therefore averages over many dynamical timescales, $\tdyn = \OmegaK^{-1}$,
and correspond to time/space average geometrical/physical conditions in
the region where the line is produced. 
Studies of time variability of the line can add a new dimension to inferences
based on averaged spectra, as they can potentially resolve the structure 
of the X--rays production region. Reverberation analysis (investigations
of the response of the line to variability of the continuum driving its
emission; Blandford \& McKee 1982) can provide independent constraints on 
the geometry of the emission region (e.g.\ Reynolds et al.\ 1999). 

Indeed, a number of surprising results were obtained in those cases where 
a good quality of data enabled studies of the line variability. Generally, 
the line absolute flux appears to be less variable than the flux 
of the driving continuum. The best example here is the long campaign 
of Rossi X-ray Timing Explorer ({\it RXTE\/}) observations of MCG--6-30-15 
analyzed in Reynolds (2000) and Lee et al.\ (2000). 
The r.m.s.\ variability in the line
band was reduced to $\approx 0.8\times$ that of the continuum, due to
a combination of constant line flux and flux-correlated spectral changes.
Similarly, Done et al.\ (2000) analyzed \xte observations of another 
Seyfert 1 galaxy
IC~4329a and found that the absolute amplitude of the reprocessed component
stayed roughly constant as the source brightened, i.e.\ the relative amplitude
decreased somewhat. Chiang et al.\ (2000) found constant line flux in
NGC~5548, even though the 2-10 keV continuum flux varied by a factor of 2.

However, the lack of response of the line integrated flux to the continuum
variability does not mean that there are no changes at all. The profile
of the line clearly responds to the continuum variability. 
Again, the most striking
example is MCG--6-30-15, where the line profile showed quite dramatic 
changes (Iwasawa et al.\ 1996): it was narrow during a high flux state but
it was very broad when the source's flux decreased. Interestingly,
the total flux of the line photons seems to have been constant in time
(see fig.\ 6 in Iwasawa et al.\ 1996). 

A suggestion that the line flux can be constant due to ionization of the
reprocessor varying in relation to the primary flux were made by Reynolds 
(2000). However, no physical model was proposed for the flux--ionization
dependence. Nayakshin, Kazanas \& Kallman (2000; hereafter NKK) pointed out 
that a reduction of the Fe \Ka line
variability can be expected if the disc develops a hot, ionized layer where
the line production is inefficient. Development of such an ionized layer is
a necessary consequence of the thermal instability operating in X--ray
irradiated plasma (Krolik, McKee \& Tarter 1981;
R\'{o}\.{z}a\'{n}ska \& Czerny 1996, hereafter RC96). 
Since the thickness of the
hot layer and its radial extent are correlated with the illuminating 
flux, the efficiency of the \Ka line production can indeed be anti-correlated
with the source X--ray luminosity.
Moreover, changes to the radial location of the line emission region
will influence the line profile.

The same idea has been proposed by Nayakshin (2000a,b) to explain the X-ray 
``Baldwin effect'' (Baldwin 1977; Iwasawa \& Taniguchi 1993; Nandra et al.\ 
1997), 
i.e.\ smaller equivalent width of the \Ka line for X-ray brighter 
sources. The two effects (the Baldwin effect and the lack of reverberation
signatures) can be simply considered as the 
same effect looked at statistically in a sample of objects and in a time 
history of a single object, respectively (``ergodic theorem''), although
the two phenomena may differ in details: X--ray variability of a single 
source is probably related to variable
output of some magnetic activity, while different (averaged) luminosities
of sources are probably due to different mass accretion rates.

In this paper we consider and test this model further, based on computations 
of vertical  structure of illuminated  accretion discs in hydrostatic 
equilibrium  (RC96;  R\'{o}\.{z}a\'{n}ska 1999, hereafter R99; see also NKK).
The plan of the paper is as follows: in Section~\ref{sec:phys} we sketch 
the physics of the thermal instability responsible for the two-layer structure
of irradiated accretion discs, in Section~\ref{sec:thickness} we present
our method of computing the thickness of the hot layer and results of
such computations, Section~\ref{sec:timing} contains discussion of 
consequences of the presence of the hot layer for variability properties
of accreting objects, while Section~\ref{sec:specdata} contains similar
discussion on spectral properties. The results are discussed 
in  Section~\ref{sec:discus}.

\section{Physical picture}
\label{sec:phys}

\subsection{Thermal instability of irradiated plasma}

Our physical picture is based on vertical structure of illuminated accretion
discs. As a result of illumination the topmost
disc layer can be strongly ionized, but the level of ionization decreases
with depth into the disc. 
In calculations performed with the assumption of constant density
(Ross \& Fabian 1993; \.{Z}ycki et al.\ 1994), 
the ionization level at the disc surface is given by $\xi\equiv \FX/\ne$
($\FX$ is the flux of illuminating X-rays and $\ne$ is the electron number 
density) and it smoothly decreases inwards. Adjusting any of these parameters 
it is possible to obtain any specified ionization stage.
The picture is rather different when the condition of 
hydrostatic equilibrium (or in fact any condition on gas pressure rather 
than its density) is additionally imposed in the illuminated disc
(RC96; NKK). Solutions
corresponding to certain range of temperatures (and/or ionization
parameter $\Xi \equiv \FX/(c\Pgas)$, more appropriate than $\xi$ for 
characterizing solutions with pressure constraints) are thermally unstable
(Field 1965; Krolik et al.\ 1981). The condition for instability
(Field 1965), 
\begin{equation}
\left( {\partial {\cal L} \over \partial T } \right)_{\rm P} < 0,
\end{equation}
where $\cal L$ is the heat-loss function, can be expressed as
\begin{equation}
\left( {\d T \over \d \Xi} \right) < 0.
\end{equation}
The unstable solutions are thus those located on the part of the
$T$--$\Xi$ diagram with a negative slope (Krolik et al.\ 1981).
The physical mechanism for the instability is the weak dependence
of bremsstrahlung cooling on temperature,
$j_{\rm brems} \propto T^{1/2} $, which cannot balance 
the photoionization heating on heavy elements caused by hard X--rays. 
On the lower stable branch at $T \sim 10^4$ K
the plasma is cooled by atomic line emission, while on the upper branch, at
$T\sim 10^7$--$10^8$ K, Compton heating and cooling are the dominant
processes.

\subsection{Vertical structure of irradiated discs}

In an illuminated accretion disc in hydrostatic equilibrium 
a movement along the $T$--$\Xi$ curve
corresponds to changing the depth into the disc.
As a result of the instability, a broad range of intermediate ionization
of iron is inaccessible to a stable solution. 
The uniqueness of the solution is assured only when the basic equations:
energy equation, hydrostatic equilibrium and radiative transfer equations
are supplemented by the heat conduction equation (Macio{\l}ek-Nied\'{z}wiecki,
Krolik \& Zdziarski 1997; R99).
The transition from the hot to cold solution branch is very sharp 
and its location can be obtained from the general condition of 
radiative--conductive equilibrium (R\'{o}\.{z}a\'{n}ska 2000).
The condition gives the temperature at the transition point 
\begin{equation}
\Ttrans={7\over 15}\Ts,
\end{equation} 
where $\Ts$ is the surface temperature. For purely radiation
heated disc atmosphere $\Ts = \TIC$, where $\TIC$ is the inverse-Compton 
temperature (McKee \& Begelman 1990; R\'{o}\.{z}a\'{n}ska \& Czerny 2000a). 
When the $\alpha P$ heating is also included, $\Ts$ is somewhat
higher than $\TIC$ (R\'{o}\.{z}a\'{n}ska 2000).
In a good approximation, the transition
occurs where the upper, stable branch of the $\Xi$--$T$ curve changes to
the middle, unstable branch (RC96, R99, NKK), 
The resulting structure is then basically two-layered: the upper, hot layer 
(HL)  and the lower, cold layer, although there is a thin transition layer
in between the two (R99). 
For a sufficiently hard illuminating spectra (i.e.\ high $\TIC$) iron can 
be considered completely ionized in the hot layer, while it recombines to 
below Fe{\sc XI} in the cold layer
(NKK; see also fig.\ 2 in \.{Z}ycki \& Czerny 1994).

\section{The thickness of the hot layer}
\label{sec:thickness}

\begin{figure*}
\epsfysize = 6cm
\hfil\epsfbox[10 500 600 700]{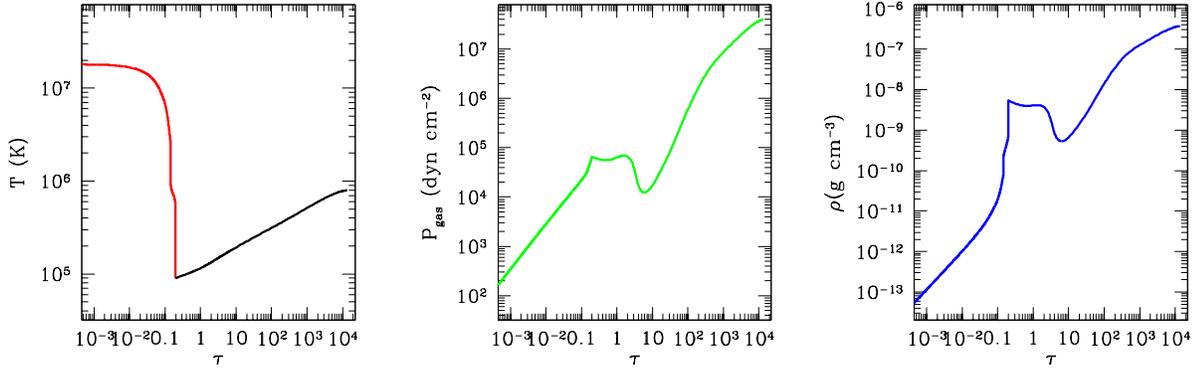}\hfil
\caption{
Vertical structure of the hot layer on top of an accretion disc
at $r=10\,\Rg$, for $M=10^7\,\MSun$, $\dot m = 0.05$, $\alpha=0.1$,
$\xi = 0.5$ and $\eta=1$.
Plotted are the gas temperature, gas pressure and density. Sharp transition
between the upper, hot, ionized layer and cool disc is observed. The thickness
of the hot layer (from $\Ts\approx \TIC$ to $7/15\,\Ts$) 
is $\tauh \approx 0.1$. The non-monotonicity
of $\Pgas$ and $\rho$ is due to contribution to energy transfer from 
vertical convection, and contribution from radiation pressure.
\label{fig:vert1}}
\end{figure*}

The model parameter crucial for spectral/timing predictions is the thickness 
of the hot layer, $\tauh$. The larger the thickness, 
the smaller the
fraction of primary radiation penetrating to the cold layer and giving 
rise to the usual reprocessed component. The reprocessed
component is further Comptonized as the photons escape through the hot layer.
The result (at least in the limited energy band corresponding to e.g.\ 
{\it Ginga\/} or \xte data)
is a reprocessed component with amplitude reduced compared to the usual
geometrical factor $\Omega/4\pi$ (Eq.~\ref{equ:refl}), 
where $\Omega$ is the solid angle
subtended by the reprocessor from the X--ray source (Done \& Nayakshin 2001).
The amplitude of the iron spectral features is further reduced by 
the Comptonization.

Two most important parameters determining the thickness of the hot layer 
are: the strength of irradiation compared to internal disc emission, $\FX/\FD$,
and the ratio of gravity at the base of the layer to X--ray radiation pressure
(RC96, NKK).
Following NKK we parameterize the latter by 
\begin{equation}
\label{equ:apar}
A\equiv {\mu_{\rm m} \over \me} {\Rg\over r} \left( {\Hd \over r}\right)^2 
 \lx^{-1},
\end{equation}
where
\begin{equation}
\lx \equiv {\FX \Hd \sigmaT \over \me c^3},
\end{equation}
where $\mu_{\rm m} = \rho/\nH \approx 1.9$ for cosmic abundance.
Obviously then, the structure of the hot layer has to be solved simultaneously 
with the structure of the cold disc, since $A$ depends on the disc thickness,
$\Hd$. NKK used vertically averaged solutions of Shakura \& Sunyaev (1973;
hereafter SS)
for the disc in their computations of
$\tauh$. They assumed that total pressure, $\Pgas+\Prad$, should be continuous
 across the 
boundary between the two layers, and obtained both $\Pgas$ and $\Prad$ 
individually continuous across the boundary.

In this paper we explicitly solve  equations of vertical structure of the cold
disc. We do this for the $\alpha\Ptot$ SS disc, i.e.\ a disc with energy 
generation proportional to total pressure. We define the input parameters
of the computations in such a way as to make the connection between accretion
discs models and observable quantities obvious. First, we 
describe calculations at a given radius and then discuss radial
dependence of $\tauh$.

\subsection{Vertical structure at a given radius}
\label{sec:vert}

The structure of the HL at a given radius is computed by combining
the method of RC96 and R99 with computations
of the vertical structure of X-ray illuminated discs by R\'{o}\.{z}a\'{n}ska
et al.\ (1999). For the HL we follow closely the method of RC96 and R99,
with the important simplification of neglecting thermal conduction. 
The reason is purely practical, as efforts to fully combine proper 
photo-ionization computations of the hot layer with vertical disc structure 
have only just begun (Dumont, Abrassart \& Collin 2000; R\'{o}\.{z}a\'{n}ska
et al., in preparation). This leaves us with an important problem
of selecting proper solution in the zone of instability, where the system 
of equations has more than  one solution for temperature. We adopt a simple
prescription and select the highest value of $T$ as the
solution. Comparing solutions with and without conduction in RC96 and R99
one sees that our procedure may overestimate somewhat the total thickness of 
the hot and transition layers (i.e.\ the depth of the point where $T=\Teff$).
However, the thickness of the hot layer alone is not affected by our
neglecting the thermal conduction.
The spectrum of illuminating radiation is assumed to be a power law with 
a cutoff,
$F_E \propto E^{-\Gamma+1} {\rm e}^{-E/E_{\rm c}}$, parameterized by
the photon spectral index $\Gamma$ and cutoff energy, $E_{\rm c}$.

Equations of vertical structure of the cold disc are the same as in
R\'{o}\.{z}a\'{n}ska et al.\ (1999). We assume that a certain fraction
of gravitational energy, $\xi$, is dissipated within the disc  (but the disc 
transports all the angular momentum, see e.g.\ Witt, Czerny \& \.{Z}ycki 1997).
The remaining fraction, $1-\xi$, is dissipated in an active corona and 
converted to hard X-ray radiation, 
of which a fraction $0.5 \eta$ illuminates 
the disk, i.e.
\begin{equation}
\FX  = 0.5 \eta (1-\xi) \Ftot,
\end{equation}
where 
\begin{equation}
\label{equ:gravity}
\Ftot = G(r) \equiv {3 \over 8\pi} {G M {\dot M} \over r^3} 
 \left(1-\sqrt{{6\Rg\over r}}\right),
\end{equation}
is the gravitational energy dissipation per unit area of the disc surface.
Value of $\eta=1$ would
thus correspond to a stationary illumination by an X-ray producing corona  
above the disk, assuming isotropic emission 
(see e.g.\ Haardt \& Maraschi 1991).
We will assume $\eta \ge 1$, corresponding to a possible enhancement of the 
X--ray flux. Such an enhancement can be expected if the X--rays are produced by
magnetic flares, due to 1) the ``charge'' time of a flare being longer
than the ``discharge'' time, 2) the collecting area for the flare energy
being larger than the illuminated area (e.g.\ Haardt, Maraschi \& Ghisellini 
1994). 

We solve the structure of the HL assuming a certain initial geometrical 
thickness of the total system, $\Htot$. Four parameters, computed
at the bottom of the HL, are then used to modify the equations of the disc 
structure and their boundary  conditions:
\begin{itemize}
 \item Location of the bottom of the HL (height above the equatorial plane)
 gives the  initial thickness  of   the cold disc,
 \item The hard X-ray flux attenuated by the HL contributes
  to heating of the disc.
 \item The temperature at the top of the cold disc is computed from 
 the Eddington approximation, 
 $T=\Teff[1/2+(3/4)\,\tau]^{1/4}$, taking $\tau$ to be  equal to 
 the optical depth of the HL, and $\Teff$ from both the internal energy 
  dissipation  and the irradiation, 
 \item Gas pressure at the bottom of the HL determines the gas
  density at the top of the disk.
\end{itemize}

Iterations of the vertical structure of the disc now follow, to compute
the disc thickness for given disc dissipation $\xi \Ftot$, 
with the boundary conditions as above.
Usually, the converged thickness of the disc is different than the initially
assumed one. Therefore the final solution (at a given radius $r$ and for an 
assumed central mass $M$, mass accretion rate $\dot m \equiv {\dot M}/\MEdd$, 
and viscosity parameter, $\alpha$)  is obtained by iterating $\Htot$
until a consistent solution is found. As already mentioned above, we define
$\tauh$ as the thickness of the layer
where temperature changes from $\Ts$ down to $(7/15)\,\Ts$.

As a reference, we compute $\tauh$ for parameters appropriate for a
Seyfert 1 galaxy:
$M=10^7\,\MSun$, 
$\dot m = 0.05$, 
$r=10\,\Rg$ (close to the peak in energy dissipation),
$\alpha = 0.1$,
$\xi=0.5$ and $\eta=1$ (i.e.\ no enhancement),
$\Gamma=1.9$ and $E_{\rm c} = 150$ keV.

Figure~\ref{fig:vert1} shows an example of the vertical structure of the
disc with the hot layer. The Thomson thickness of HL is rather small,  
$\tauh\approx 0.1$, irrespectively of $\alpha$. The ``gravity parameter''
is $A\approx 7$ in this case. Our value of $\tauh$ is thus very similar
to what NKK obtain for this value of $A$ (see fig.~4 in NKK), demonstrating
that, despite some approximations in our procedure, our results are
accurate.

\begin{figure*}
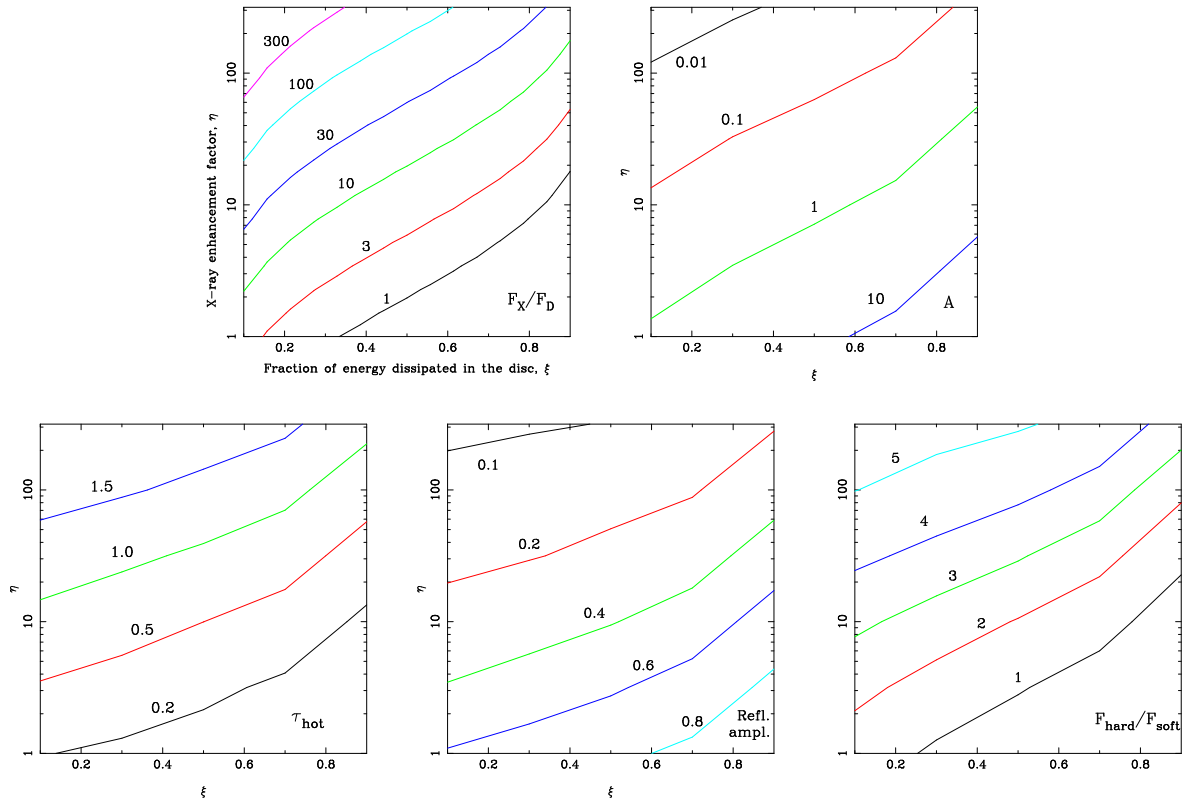

 \parbox{\textwidth}{ 
 \hfil
  \parbox{0.48\textwidth}{ 
    \epsfxsize = 0.3\textwidth
    \epsfbox[0 150 550 740]{Fx2Fd_r5Rs_m0x05.ps}} 
  \parbox{0.48\textwidth}{ 
    \epsfxsize = 0.3\textwidth
    \epsfbox[0 150 550 740]{Apar_r5Rs_m0x05.ps}}
 \hfil}
 \parbox{\textwidth}{ 
 \hfil\parbox{0.31\textwidth}{ 
  \epsfxsize = 0.3\textwidth
  \epsfbox[0 150 550 720]{Tauhot_r5Rs_m0x05.ps}} 
\parbox{0.31\textwidth}{ 
  \epsfxsize = 0.3\textwidth
  \epsfbox[0 150 550 720]{Refl_r5Rs_m0x05.ps}}
\parbox{0.31\textwidth}{ 
  \epsfxsize = 0.3\textwidth
  \epsfbox[0 150 550 720]{Fh2Fs_r5Rs_m0x05.ps}}
 \hfil}
 \caption{
Results of computations of the structure of the hot layer on top of an 
accretion disc at $r=10\,\Rg$, for $M=10^7\,\MSun$, $\dot m=0.05$ 
and viscosity parameter  $\alpha=0.1$. The panels show contours of: 
$\FX/\FD$ -- the ratio of the illuminating X--ray flux to the disc flux due
to internal dissipation; 
$A$ -- the gravity parameter (see equation~\ref{equ:apar});
$\tauh$ -- the Thomson thickness of the hot layer; 
the effective amplitude of the cold, reprocessed component 
(equation~\ref{equ:refl});
$\Fhard/\Fsoft$ -- the ratio of the observed hard
X--ray flux to the soft, thermal flux.
\label{fig:cont1}}
\end{figure*}

The main parameter, $A$, is a non-unique function of other parameters of 
the system. For given $\dot m$ and $r$, $A$ is determined
by a combination of $\xi$ and $\eta$: both parameters influence
the disc thickness and the irradiation flux.
We have therefore extended our reference computations
to obtain $\tauh$ on a grid of $\xi$ and $\eta$. Results are plotted in
Figure~\ref{fig:cont1}, where we show a number of quantities characterizing
the disc or the spectrum. 

The amplitude of the observed ``cold'' reflection (energy-integrated albedo)
is estimated using
\begin{equation}
\label{equ:refl}
R = {\left[\Ptrans(\tauh)\right]^2 \over 1+\Prefl(\tauh)},
\end{equation}
where $\Ptrans(\tauh)$ and $\Prefl(\tauh)$ are hard X--rays transmission and 
reflection probabilities for the HL thickness $\tauh$. 
They are defined as ratios of the transmitted and reflected flux to the 
incident flux, respectively, for a purely scattering atmosphere.
The factor $\Ptrans^2$ appears in the numerator of equation~(\ref{equ:refl}), 
because the reflected photons have to go through the HL {\em twice}.
Additionally, a fraction $\Prefl$ of the incident photons is elastically 
back-scattered off the HL, and, by our assumption, simply  contributes to
the primary continuum photons initially directed towards an observer.
In the limiting case of no HL ($\tauh \rightarrow 0$), 
$\Ptrans(\tauh) \rightarrow 1$, while $\Prefl(\tauh)\rightarrow 0$ and
$R\rightarrow 1$, i.e.\ $R$ represents then the solid angle of the
reprocessor, $\Omega$, normalized to $2\pi$.
We compute $\Ptrans(\tauh)$ and $\Prefl(\tauh)$ using a 
Monte Carlo simulations of Compton scattering.
This is a rather approximate estimate of the angle averaged
amplitude, since it assumes no photo-absorption in the HL, i.e.\ no 
contribution to the reprocessed spectrum from highly (but not completely)
ionized elements. It thus assumes a rather hard illuminating spectrum 
($\Gamma < 2$) giving high $\TIC$. A more accurate procedure of estimating
$R$ was employed by Done \& Nayakshin (2001), who fitted spectra computed
by NKK by a ``cold'' reflection model in the context of {\it Ginga\/} data.
For softer illuminating spectra, $\Gamma \ga 2$, the reprocessed
component contains strong spectral features due to H- and He-like Fe ions
and it cannot be approximated by ``cold'' reprocessing (NKK; Nayakshin 2000b).

The ratio $\Fhard/\Fsoft$ is the observed ratio of fluxes, i.e.\ the hard
X--ray flux 
\begin{equation}
\label{equ:fhard}
\Fhard = \FX (1+\Prefl)
\end{equation}
contains a contribution from photons backscattered in the HL  towards 
an observer. The soft flux, $\Fsoft$, contains a contribution from the disc
internal emission, $\FD = \xi \Ftot$, and hard photons
thermalized in the cold disc, 
\begin{equation}
\label{equ:fsoft}
\Fsoft = \FD + (1-a) \Ptrans \FX,
\end{equation}
with $a\approx 0.2$ being the energy-integrated X--ray albedo of cold matter.
We note that the above relations of $\Fhard$ and $\Fsoft$ with $\FX$
assume implicitly that there are no significant changes of the source geometry
(e.g.\ the height of the flare above the disc), so that the variability of
$\FX$ is solely due to the variable output of the magnetic activity.

The structure of the disc has to be rather different from the standard SS 
solution, if the hot layer is to be Thomson thick(ish), $\tauh \ga 1$.
For half of the total energy dissipated in the active
regions, $\xi=0.5$, the illumination flux has to be further enhanced by a 
factor of $\eta\approx 40$ relative to the stationary situation
in order to obtain $\tauh=1$.
This corresponds to the ratio of illumination flux to disc internal flux,
$\FX/\FD =0.5 \eta (1-\xi)/\xi \approx 20$. 
For larger fraction of energy dissipated in the
magnetic corona, i.e.\ smaller $\xi$, the required ratio $\FX/\FD$ is even 
larger (e.g.\ $\FX/\FD\approx 60$ for $\xi=0.1$). This appears to be
due to the disc thickness, $\Hd$, increasing with $\xi$ more slowly
than linearly, and thus for $A\propto (\FD/\FX) (\Hd/\xi)$ to remain
constant, $\FX/\FD$ has to increase when $\xi$ decreases.

On the other hand, the presence of the HL will be apparent in observations
even if its thickness is rather smaller than $\tauh=1$. According to
our approximate formula (equation~\ref{equ:refl}) reflection amplitude 
gets reduced to 0.8 for $\tauh \approx 0.25$. Quality of current data is 
usually sufficient to distinguish $R=0.8$ from $R=1$ in typical X--ray spectra
of AGN (e.g.\ Done et al.\ 2000).

\subsection{Radial structure}
\label{sec:rad1}

\begin{figure}
 \epsfxsize = 0.45\textwidth
 \epsfbox[20 300 550 700]{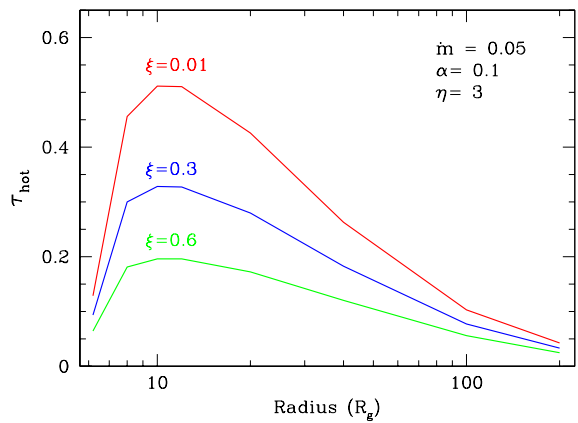}
 \caption{
Thomson thickness  of the hot layer as a function of radius for
$\eta=3$ and three values of $\xi$, as labelled.
Other parameters: $M=10^7\,\MSun$, $\dot m=0.05$ and $\alpha=0.1$. 
\label{fig:taurad}}
\end{figure}

The Thomson thickness of the HL changes with radius, even for
a constant $\FX/\FD$, since $A$ is not constant with $r$. 
We computed radial dependencies of $\tauh$ for $\eta=3$ and three values
of $\xi = 0.01, 0.3$ and $0.6$. Values of other parameters were as before: 
$M = 10^7\,\MSun$, $\dot m = 0.05$, $\alpha=0.1$. Results are plotted
in Fig.~\ref{fig:taurad}.

The thickness is largest at $r\approx 10\,\Rg$. It then drops slowly,
roughly in accord with the formula of Nayakshin (2000a).
A consequence of that particular $\tauh(r)$ distribution is a relative 
suppression of highly relativistically smeared component of the reprocessed
spectrum. On the other hand, the reprocessed photons that do manage to
diffuse from below the hot layer close to the central BH, suffer Comptonization
smearing, which in practice may be difficult to distinguish from
relativistic smearing (see Sec.~\ref{sec:rin_vs_r}).

Obviously, in the magnetic flare scenario there is no reason for changes
of $\xi$ and $\eta$ to be perfectly correlated on all radii. One can thus
imagine rather different values at different radii, leading to $\tauh(r)$
rather different than those plotted in Fig.~\ref{fig:taurad}. 
On timescales long
compared to $\tdyn$ one could perhaps obtain time average $\xi(r)$ and
$\eta(r)$ employing a particular variability model, e.g.\ that of
Poutanen \& Fabian (1999).

\subsection{Dependence of ${\boldmath \tau_{\bf hot}}$ on X--ray flux}
\label{sec:tauflux}

The qualitative relation between the thickness of the HL and the illuminating
X--ray flux is obvious: the stronger the X--ray flux the thicker the HL.
In this Section we present the quantitative relation from our model.

\begin{figure}
 \epsfxsize = 0.45\textwidth
 \epsfbox[0 150 550 720]{refl_fhard_tmp.ps}
 \epsfxsize = 0.45\textwidth
 \epsfbox[40 500 550 700]{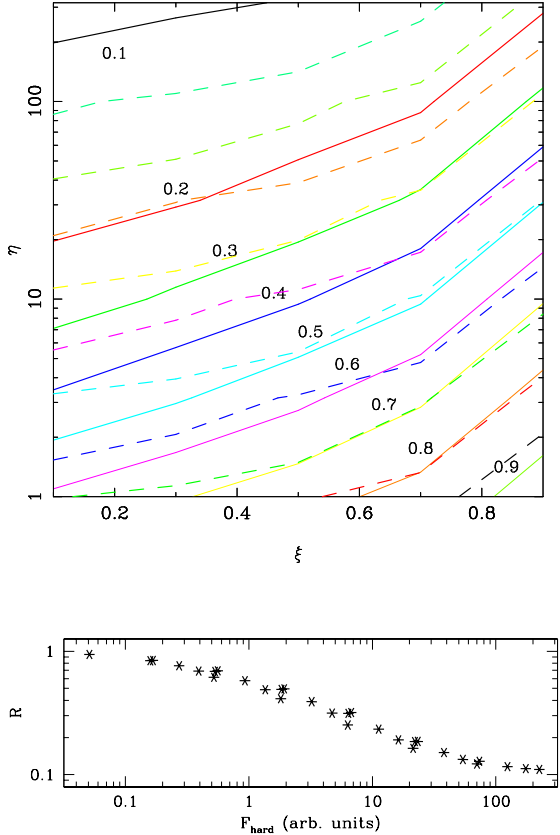}
 \caption{
{\it Upper panel:\/} contours of observed amplitude of the reprocessed 
 component (solid curves, 
labelled), and contours of observed hard X--ray flux (dashed curves).
Each contour of $\Fhard$ means an increase of $\Fhard$ by a factor of 2.
{\it Lower panel:\/} projected dependence of $R$ vs.\ $\Fhard$. Clearly,
the dependence is much weaker than linear, i.e.\ it is {\it not\/} possible
for the absolute normalization of the reprocessed component, 
 $N = R\times \Fhard$ to remain constant. See Figure~\ref{fig:cont1} for
parameters of computations.
\label{fig:reflfhard}}
\end{figure}

In Figure~\ref{fig:reflfhard} we again plot  contours of amplitude
of reflection $R$ in the $\xi$--$\eta$ plane, together with contours of 
the observed X--ray flux, $\Fhard$, as well as the projected dependence 
$R$--$\Fhard$. 
For $R$ close to 1, doubling $\FX$ results in only
modest decrease of $R$, by 10--20 per cent. For smaller $R$ the process
is more effective. But still, to go from e.g.\ $R=0.8$ to $R=0.4$ requires 
an increase of $\Fhard$ by a factor of $\ga 8$ (when the movement is 
along the $\xi$ axis, for $\eta \approx 4$). Going from $R=0.6$ to $R=0.3$
requires that $\Fhard$ increases at least 4 times, again approximately
along the $\xi$ axis. In no case does it seems possible to obtain an
inverse proportionality of $\Fhard$ and $R$, so that 
$R\times \Fhard \approx {\rm const}$. Thus, these results seem to rule out the
possibility that the absolute flux of the Fe \Ka line may remain
constant despite continuum variability, although certain reduction of 
the line variability amplitude (compared to that of the continuum)
can be expected (see next Section).

We note that a change of $\FX$ is more effective in changing $\tauh$,
if $\FX$ increases or decreases due to changing fraction of energy dissipated
in the disc, $\xi$, rather than $\eta$. The reason is 
that e.g.\ removing the energy dissipation from the disc to the active corona
(i.e.\ decreasing $\xi$) not only increases the X--ray illuminating flux,
but it also decreases the disc thickness, thus decreasing the gravity.

\section{Reduction of variability amplitude of the Fe 
K$_{\boldmath \alpha}$ line}
\label{sec:timing}

\subsection{Method}
\label{sec:reducmeth}

Based on results of previous Section we expect the variability of the 
Fe \Ka line flux to be reduced compared to the amplitude
of variability of the driving continuum, even though it does not seem possible 
to obtain absolutely constant flux of the line. 
Here we estimate the reduction of the r.m.s.\ variability of the Fe \Ka
line, compared with an assumed r.m.s.\ variability of its driving
continuum. To this end we will use results of our calculations of
$\tauh(\xi, \eta)$ at $r=10\,\Rg$. In this way we will obtain the upper
limit to the considered effect, since $\tauh(r)$ reaches maximum at 
$r\approx 10\,\Rg$, for a given pair $(\xi, \eta)$ (see Fig.~\ref{fig:taurad}).

First, we construct a simulated light curve of hard X--ray continuum by summing
Fourier components with random phases (e.g.\ Tsonis 1992),
\begin{equation}
 C(t_i) = \sum_{j=1}^{N} \sqrt{P(f_j)} \cos\left[2\pi f_j t_i + 
   \phi(f_j)\right],
\end{equation}
for an adopted form of power spectral density (PSD), $P(f)$.
For the PSD we adopt
\begin{equation}
 P(f) \propto {1 \over \left[1+(f/\fc)\right]^p}.
\end{equation}
This provides a good description of hard X--ray PSD of e.g.\ MCG--6-30-15, with
the cutoff frequency $\fc=10^{-5}$ Hz, the slope $p=1.4$ 
and the normalization corresponding to r.m.s.\ variability of 20 per cent
(Green, McHardy \& Lehto 1993; Czerny \& Lehto 1997; Yaqoob et al.\ 1997; 
Nowak \& Chiang 2000; Reynolds 2000).
We take these parameters to be representative for Seyfert 1 galaxies.
The light curve is scaled to have a given mean value ${\overline C}$.
Next, the light curve in energy band containing the \Ka line photons
is constructed according to
\begin{equation}
\label{equ:lineflux}
L(t_i) = (1-\Al) C(t_i) + \Al C(t_i) R[C(t_i)],
\end{equation}
where $\Al$ is the fraction of line photons contributing to 
total counts in the considered energy band, while $R[C(t)]$ is the relative 
amplitude of the reprocessed component as a function of the illuminating 
hard X--ray flux. 
Writing equation~(\ref{equ:lineflux}) we have assumed no time lag in the
response of the Fe \Ka line emission after the continuum variations,
consistent with observations (e.g.\ Reynolds 2000). 
The amplitude $R[C(t)]$ is computed as follows:
we choose a value for the effective  amplitude of the reprocessed component
at the mean flux level, $R({\overline C})$. 
We then assume that $R({\overline C})$ corresponds to
a pair $(\xi_0, \eta_0)$, which is located on a trajectory in the $\xi$--$\eta$
plane that the system travels while changing its luminosity. 
The trajectory describes how changes of $\xi$ and $\eta$ contribute
to the variability. Since our understanding of variability mechanisms
is  rather limited, we will simply test a number of arbitrary trajectories,
to get an idea of available range of results.  As discussed in previous 
Section, variations along the $\xi$-axis give largest possible change of $R$ 
for a given change of $\FX$, therefore our chosen trajectories have a 
significant component along the $\xi$ axis.

Next, we construct the light curve of the observed
hard X--ray continuum, which has a contribution from photons backscattered
in the hot layer, i.e.\ $F(t) = C(t) \{1+\Prefl\{\tauh[C(t)]\}\}$. 
Finally, the
r.m.s.\ variability is computed for $F(t)$ and $L(t)$ according to
\begin{equation}
\sigvar(C) = {1 \over {\overline C}}{\sqrt{{1\over N} 
\sum_{i=1}^{N}\,\left[C(t_i)-{\overline C}\right]^2}}.
\end{equation}

\begin{figure}
 \epsfxsize = 0.45\textwidth
 \epsfbox[0 250 550 720]{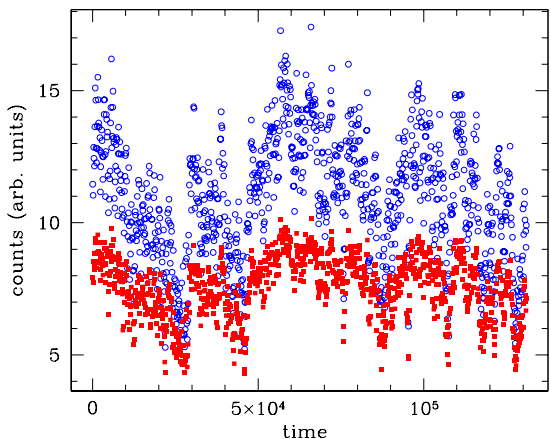}
 \caption{Examples of light curves of the observed hard X--ray continuum 
  (open circles)  and the Fe \Ka line (solid squares), simulated from the PSD
  defined in Sec~\ref{sec:reducmeth}. The assumed parameters gave 
  the effective amplitude of the reprocessed component 
  $R({\overline C})\approx 0.75$. In effect the line flux is suppressed
  at almost all times and reduction of the line r.m.s.\ variability is 
  significant,  at 6 per cent   (see Fig.~\ref{fig:reductvar}).
\label{fig:simlcurve}}
\end{figure}

\subsection{Results}
\label{sec:reductvar}

\begin{figure}
 \epsfxsize = 0.45\textwidth
 \epsfbox[0 150 550 720]{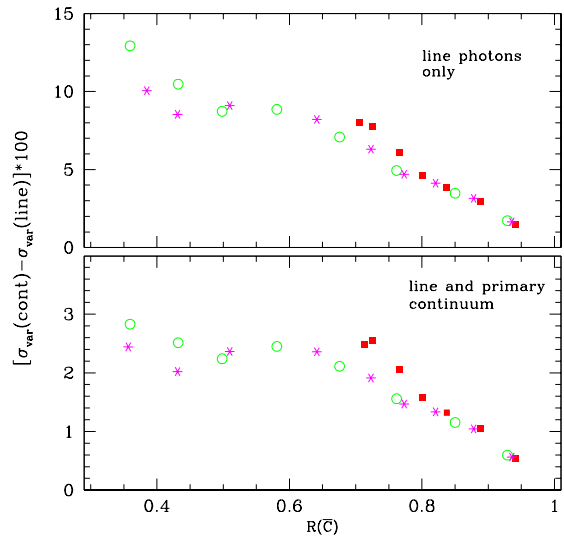}
 \caption{Amplitude of variability of the Fe \Ka line flux is reduced compared
 the that of the driving continuum, due to appearance of a hot, ionized layer
 with thickness increasing with increasing illuminating flux. This is plotted
 here as a function of the assumed effective amplitude of reflection  
 at the mean   flux level.
 {\it Upper panel\/} shows reduction of the r.m.s.\ variability assuming that
  there are no primary continuum photons in the line energy band.
 {\it Lower panel\/} shows analogous reduction assuming that the line photons
  contribute 15 per cent to the considered energy band.
  Different symbols denote different trajectories in the $\xi$--$\eta$ plane
  traveled by the varying system. The r.m.s.\ variability of the driving
  continuum was 20 per cent. See text  (Sec~\ref{sec:reductvar}) for 
  further explanations.
\label{fig:reductvar}}
\end{figure}

First we set $\Al=1$ i.e.\ we consider the light curve of the Fe \Ka line
alone, without any continuum photons. An example of the simulated light curves
is plotted in Fig.~\ref{fig:simlcurve}. They were obtained assuming that
the mean flux of the intrinsic continuum, $\overline C$, corresponds to 
$\xi_0 = 0.5$ and $\eta_0 = 1$ (which yields $R({\overline C})\approx 0.75$,
see Fig.~\ref{fig:reflfhard}),
and that the variability is only due to changing $\xi$. The line flux is
reduced at all times, since $R({\overline C})$ is significantly lower
than 1, and its r.m.s.\ variability is reduced by $\approx 5$ per cent
compared with the r.m.s.\ variability of its driving continuum, $C(t)$. 
The observed hard X--ray continuum, $F(t)$, contains a contribution from 
photons
backscattered in the hot layer, which increase its r.m.s.\ variability
by $\approx 1$ per cent. Thus, the difference of r.m.s.\ variabilities
between the line and the hard continuum, $\sigvar(F)-\sigvar(L)$ is 
in this example $\approx 6$ per cent.

In real spectra the contribution of line photons to the 5--7 keV band
is only $\Al\approx 0.15$, for $R=1$. The reduction of r.m.s.\ variability,
$\Delta\sigvar \equiv \sigvar(F)-\sigvar(L)$, is
therefore much weaker than in previous case, and in the above example,
only $\approx 2$ per cent.

\begin{figure*}
 \epsfysize = 7 cm
\hfil \epsfbox[20 420 620 700]{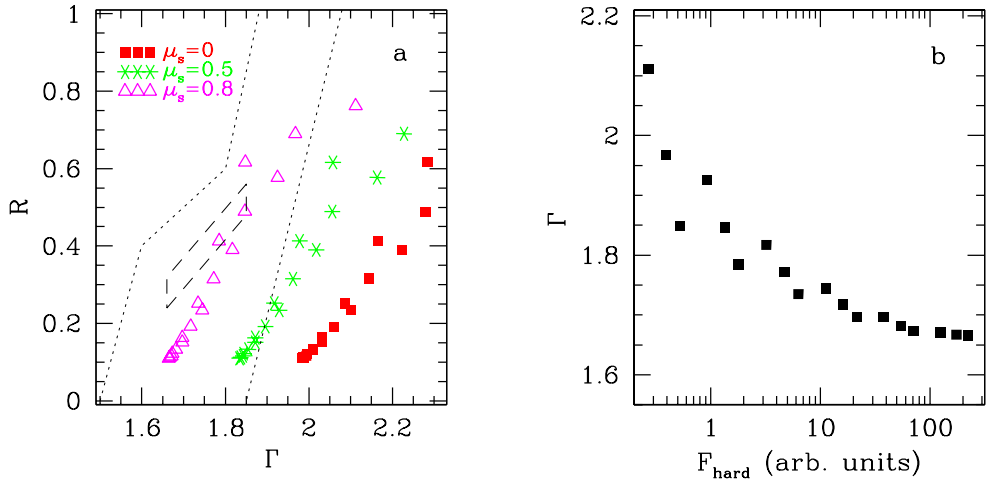} \hfil
 \caption{ Panel a): the amplitude of reflection -- spectral index 
 correlation predicted  by the model for a number of $\mus$ values 
 (equation~\ref{equ:mus}). 
 Only models with $\xi \le 0.5$ were used (most of the energy dissipated
 in the active corona).
 The contours represent the observed correlations (Zdziarski et al.\ 1999; 
 Gilfanov et al.\  2000). Panel b) shows hard X--ray flux -- spectral
 index anti-correlation. Each ``triplet'' of points corresponds to a
 given value of $\eta$ and $\xi=0.1,\ 0.3,\ 0.5$. The anti-correlation, which
 has an obvious physical interpretation in this model, seems to be opposite
 to what is observed in the data (e.g.\ Done et al.\ 2000).
\label{fig:r_vs_gamma}}
\end{figure*}

The difference $\Delta\sigvar$ is plotted in Fig.~\ref{fig:reductvar}
as a function of $R({\overline C})$, for both $\Al=1$ and $\Al=0.15$.
The general trend is that the larger the $R({\overline C})$ is, the
smaller the $\sigvar(F)-\sigvar(L)$ is. Consider for example 
$R({\overline C}) \approx 1$, i.e.\ at the mean flux level the 
irradiation flux is relatively weak and in consequence the hot layer is 
rather thin, $\tauh \le 0.1$. Then $R$ is reduced only when the instantaneous
flux exceeds the mean flux, but $R$ saturates at 1 whenever
$C(t) < {\overline C}$. This then gives smaller reduction of r.m.s.\ 
variability compared to a case $R({\overline C}) \sim 0.5$ when
the anti-correlation between $R$ and $C(t)$ can hold through the entire
range of $C(t)$. Details of the relation $R({\overline C})$--$\Delta\sigvar$
depend on the mapping $C(t) \rightarrow (\xi, \eta)$, i.e.\ which parameter,
$\xi$, $\eta$ or both of them, cause flux variations. Solid squares
in Fig.~\ref{fig:reductvar} show the case of variations $C(t)$ caused by 
changes of $\xi$ only (for $\eta=1$). 
Reduction of the line variability is strongest
in this case, since a given variation of $C(t) = \FX$ causes largest
possible change of $R$, as already discussed in Sec.~\ref{sec:tauflux}
and shown in Fig.~\ref{fig:reflfhard}. On the other hand, the available range
of $R({\overline C})$ is rather limited.

These results are quantitatively dependent on the assumed constancy of X--ray 
source geometry (see comments following Eq.~(\ref{equ:fsoft})). The reduction
of variability, $\Delta\sigvar$, could be larger, if the geometry were 
correlated with the X--ray luminosity, changing in such a way as to make the 
illuminating flux, $\FX$, to increase more strongly than the observed flux, 
$\Fhard$. 
However, as we are going to argue in the next Section, the combined  timing 
and spectral observational result put quite tight constraints on any
geometry--luminosity dependences.

Finally, we note that both $R({\overline C})$ and $\sigvar(F)-\sigvar(L)$
are observable quantities and their relation plotted in 
Fig.~\ref{fig:reductvar} can be used to test the model.

\section{Spectral Data}
\label{sec:specdata}

\subsection{The ${\boldmath R}$--${\boldmath\Gamma}$--flux  correlations}
\label{sec:r_gamma}

The spectral index of the hard X--ray power law, $\Gamma$, and the
observed amplitude of the reprocessed component, $R$, appear to be 
correlated in the data of both AGN and Galactic Black Holes (GBH)
(Zdziarski, Lubi\'{n}ski \& Smith 1999; Gilfanov, Churazov \& Revnivtsev 2000):
the harder the spectrum (smaller $\Gamma$), the smaller the amplitude.
Two models were considered to explain the correlation: truncated cold disc
with hot central flow (e.g.\ Esin, McClintock \& Narayan 1997) 
and relativistic outflow of emitting plasma (Beloborodov 1999ab).
The model of an accretion disc with a hot layer is also able to 
qualitatively explain the correlation,
with $\tauh$ being the control parameter: the larger $\tauh$ is, the smaller
the effective $R$ and the harder the spectrum, due to reduced soft flux
from thermalization (assuming that most of the energy dissipation occurs
in the active corona). We present here the $R$--$\Gamma$
correlation predicted by the model. The estimate of $R$ is obtained
as in Sec.~\ref{sec:vert} (equation~\ref{equ:refl}), 
while for $\Gamma$ we use the formula from Beloborodov (1999ab),
\begin{equation}
 \label{equ:gamma}
 \Gamma \approx 2.33(\lh/\ls-1)^{-1/10},
\end{equation}
where $\lh$ and $\ls$ are the power in hard and soft radiation
crossing the active region, respectively. We estimate 
\begin{equation}
\label{equ:mus}
\ls = \int_{-1}^{-\mus}[ F_{\rm dissip}(\mu) + F_{\rm reproc}(\mu)]\,d\mu,
\end{equation}
where $F_{\rm dissip} \equiv \FD = \xi\Ftot$ and 
\begin{equation}
F_{\rm reproc} = (1-a)\FX = {1 \over 2}\eta(1-\xi)(1-a)\Ftot\,
               \Ptrans\left(\tauh(\xi,\eta)\right). 
\end{equation}
The second term describes 
the contribution to the soft flux from the thermalized fraction of the
illuminating X--rays.
We have also introduced the geometrical factor $\mus$ (see Beloborodov 1999ab),
describing the fraction of soft flux actually intercepted by a magnetic flare
(in computations we assume isotropic emission). The hard power is simply 
estimated as $\lh = 2\FX = \eta(1-\xi)\Ftot$.

Figure~\ref{fig:r_vs_gamma} presents results of the computations for a number
of $\mus$ values. The shape of the model correlation is somewhat different
from the correlation observed in the data. The model requires $\mus > 0.6$
in order to obtain quantitative agreement with the data, i.e.\ $1-\mus < 0.4$
of the soft flux crossing an active region. It has to be remembered however
that the observed correlation, particularly among different objects,
is more likely to be due to changes of geometry driven by e.g.\ $\dot m$ 
rather than varying output of magnetic activity, and thus parameterizing 
the correlation by our parameters $\xi$ and $\eta$ may not be sufficient.

Another correlation applicable to AGN observations may be the 
$\Fhard$--$\Gamma$ (anti)correlation plotted in Figure~\ref{fig:r_vs_gamma}b. 
It follows from combining results shown in Fig.~\ref{fig:reflfhard}
with those in Fig.~\ref{fig:r_vs_gamma}: an increase of illuminating
flux, $\FX$, gives a decrease of $R$ and a decrease of $\Gamma$.
If the observed X--ray flux, $\Fhard$, follows the illuminating flux, $\FX$,
(i.e.\ if the Eq.~(\ref{equ:fhard}) holds), 
the predicted $\Fhard$--$\Gamma$ anticorrelation is obtained.
However, an opposite effect seems to be observed in the data from \xte
observing campaigns of some AGN, where the spectrum generally softens
($\Gamma$ increases) when the observed X--ray flux goes up 
(NGC 5548, Chiang et al.\ 2000; IC~4329a, Done et al.\ 2000). At least in
IC~4329a the spectral analysis of Done et al.\ (2000) reveals that this is 
not an artifact of limited
band pass of \xte instruments. If this observational result is generic,
it would mean changes of the flares' geometry correlated with their 
luminosity, so that the relation between $\Fhard$ and $\FX$ can be reversed.
From the discussion of variability (Sec~\ref{sec:reductvar}), we need
a correlated increase of $\tauh$ with the flare luminosity. If this is to
be accompanied by a softening of the spectrum, the flare's geometry has
to change in such a way, that the intercepted soft luminosity is larger than
the increased hard luminosity ($\lh/\ls$ has to decrease; 
Eq.~(\ref{equ:gamma})), despite the decreased soft flux from 
thermalized hard X--rays.

\subsection{Model dependence of relativistic smearing and amplitude
of reflection}
\label{sec:rin_vs_r}

The hot layer is thickest at close distances from the central black hole, 
according to results of computations of $\tauh(r)$ presented in 
Sec.~\ref{sec:rad1}. The \Ka line generation is then suppressed at small
$r$, which means that the inner disc radius determined from the line profile
may be larger than $\rms=6\,\Rg$. We quantified this effect in our model,
since indeed the line  observed in
a number of sources is narrower than in the extreme case of MCG--6-30-15
(IC~4329a, Done et al.\ 2000; Cyg X-1, Done \& \.{Z}ycki 1999; GS~2023+338, 
GS~1124-68, \.{Z}ycki, Done \& Smith 1997, 1998).

\begin{figure}
 \epsfxsize = 0.45\textwidth
 \epsfbox[0 230 550 720]{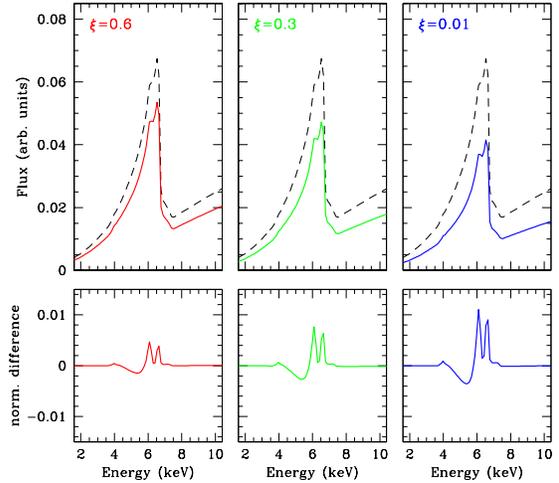}
 \caption{Reprocessed spectra computed for the radial $\tauh(r)$ 
dependencies plotted in Fig.~\ref{fig:taurad} (solid lines), compared to 
the reference case of {\em no hot skin\/} 
(full relativistic smearing; dashed line). Since $\tauh(r)$
reaches maximum at $r\approx 10\,\Rg$, the most strongly relativistically
smeared component is suppressed in the spectra. Upper panels show that
the main effect is a decrease of the overall amplitude, described by
Eq.~(\ref{equ:refl}). Lower panel show the difference,
$ionized\ skin\ spectrum - reference\ spectrum$,
(after correcting for the decrease of amplitude), normalized to the
primary power law spectrum. The additional 
Comptonization broadening of the spectral features is {\em not\/} included.
\label{fig:smearats}}
\end{figure}

We computed the relativistic smearing in the reprocessed spectra for 
the radial dependencies $\tauh(r)$ plotted in Fig.~\ref{fig:taurad}. We
assumed that the irradiation emissivity
follows radial distribution of gravitational energy dissipation,
G(r)  (equation~\ref{equ:gravity}). Ratios of the
reprocessed spectra to the reference case of no hot skin (so full relativistic
smearing), are plotted in Fig.~\ref{fig:smearats}.  The magnitude of
the effect  obviously depends on the thickness of the HL. Even for the
thickest HL considered ($\xi=0.01$, maximum $\tauh(r)\approx 0.5$),
the ``wiggles'' in the ratio plots are only a few percent, for pure
reprocessed component, i.e.\ no contribution from primary photons. 
It would then be rather difficult to detect such an effect in the data,
the more so that the additional broadening due to Comptonization in the
HL (not included here), is strongest for the thickest HL. Qualitatively then,
reduction of effective amplitude of the reprocessed component is much more
prominent, than any changes to the profile of the \Ka line (as is also clear
from Fig.~\ref{fig:smearats}).

One can estimate the maximum magnitude of the suppression of the broadest
line component, noticing that it will occur for $\tauh(r)$ being a step
function: $\tauh$ very large for radii $r<\rcut$, and $\tauh=0$ otherwise. 
In other words,
no observable reprocessed component is produced for $r < \rcut$
(Ross, Fabian \& Young 1999). This obviously
means a reduction of the effective amplitude of the reprocessed component,
which one can relate to $\rcut$

\begin{figure}
 \epsfxsize = 0.45\textwidth
 \epsfbox[0 350 550 720]{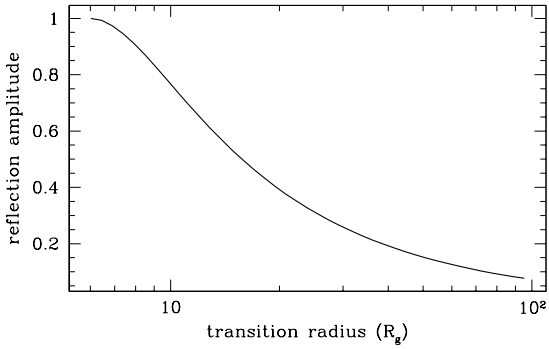}
 \caption{Relation between effective amplitude of the reprocessed component
 and the amount of relativistic smearing, for an {\em ad hoc\/} model, where
 the disc is assumed to be covered  by a perfectly reflecting HL 
(i.e.\ $\tauh \gg 1$) from $\rms$ up to a radius  $\rcut$. 
 This reduces the effective amplitude of reflection and removes
 the broadest component from the spectral features, with $\rcut$ being
 the inner disc radius as inferred from fitting the profile of the \Ka line.
\label{fig:rin_vs_r}}
\end{figure}

To do so, we write $R=\Fdown/(a_0 \Fup)$, i.e.\ $R$ is the reflection albedo
(ratio of flux of reflected radiation to the flux directed towards an 
observer), normalized to the albedo $a_0$ of cold reflection, expected
from an isotropic source above a flat disc.
Both fluxes can be written as radial integrals, e.g.\ 
\begin{equation}
\Fdown = 2\pi \int_{\rmin}^{\rmax}\Fdown(r)r\,dr.
\end{equation}
Averaged over many dynamical time scales, the radial distribution of energy 
generation in the active corona can be expected to follow the gravitational 
energy dissipation prescription.
If isotropy of emission is assumed, then 
\begin{equation}
\Fdown(r) = \Fup(r) = G(r).
\end{equation}
If a purely scattering layer of a thickness giving a transmission
probability $\Ptrans$ (reflection probability $1-\Ptrans$) covers the 
reflecting disc up to a radius $\rcut$, then
\begin{eqnarray}
\Fdown(r)/a_0 = & 2\pi \int_{\rmin}^{\rcut}\Ptrans^2 G(r) r\,dr + \nonumber \\
 &  2\pi \int_{\rcut}^{\rmax} G(r) r\,dr
\end{eqnarray}
and
\begin{eqnarray}
\Fup(r) = & 2\pi \int_{\rmin}^{\rcut} (2-\Ptrans) G(r) r\,dr + \nonumber \\
& 2\pi \int_{\rcut}^{\rmax} G(r) r\,dr.
\end{eqnarray}
The resulting function $R(\rcut)$ 
is plotted in Figure~\ref{fig:rin_vs_r} for $\Ptrans=0$, which maximizes 
$\rcut$ for any $R$. For $\Ptrans(r)=1$ (no hot layer) we obviously
obtain $R(\rcut) = 1$, but for any $\Ptrans(r) < 1$ the amplitude $R$ decreases
dramatically with increasing $\rcut$, i.e.\ when the ionized zone becomes more
extended. The reason for the dramatic decrease is that the
emission is strongly concentrated towards the center. Interestingly,
many Galactic black hole binaries show $R\approx 0.3$ and 
$\Rin\approx 20\,\Rg$, in accord with the relation shown in 
Figure~\ref{fig:rin_vs_r}, but we note that, with the assumption of perfect
reflectivity, the geometry here is equivalent to the disc being truncated
at $\rcut$, and replaced by central X--ray source.
We also emphasize that the assumption $\Ptrans(r<\rcut) = 0$, is not
justified by our present understanding of the model.

\subsection{The X--ray ``Baldwin effect''}
\label{sec:ewlum}

\begin{figure}
 \epsfxsize = 0.45\textwidth
 \epsfbox[0 200 600 700]{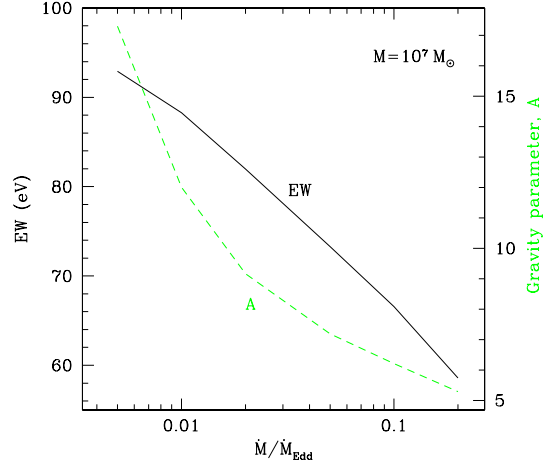}
 \caption{Equivalent width of the Fe \Ka line (solid curve, left scale) 
 as a function of the average
 X--ray source luminosity, $\LX$, assuming that $\LX \propto \dot m$, 
 i.e.\ $M={\rm const}$. EW decreases with $\dot m$ which could explain
 the observed X--ray ``Baldwin effect''.
 The dotted curve shows the corresponding gravity 
 parameter $A$, which decreases due to slower than linear increase of the
 disc thickness with $\dot m$. Parameters: $M=10^7\,\MSun$, $\alpha=0.1$,
 $\xi=0.5$, $\eta=1$.
\label{fig:ewlum}}
\end{figure}

The observed equivalent width (EW) of the Fe \Ka line appears to be 
anticorrelated with average 2--10 keV luminosities of the sources 
(Iwasawa \& Taniguchi 1993; Nandra et al.\ 1997), in a similar manner as
EW of C{\sc IV} $\lambda$1550 line is anticorrelated with the UV continuum 
luminosity (Baldwin 1977).
This could be explained by the presence of the hot layer, since indeed
$\tauh$ is larger for larger $\FX$, although in this case we expect
the flux to be driven by accretion rate, rather than $\xi$ and/or $\eta$.
Since the average X--ray luminosity can be expected to be proportional
to the accretion rate ${\dot M} = {\dot m} \MEdd \propto {\dot m} M$, 
we have computed the EW varying either
$M$ or $\dot m$ and keeping the other parameter constant. We again assumed
that situation at $r=10\,\Rg$ is  characteristic to the whole disc,
and we adopted $\xi=0.5$ and $\eta=1$.
Results  presented in Fig.~\ref{fig:ewlum} show that indeed the EW of the
\Ka line ($\propto R$) decreases with $\dot m$. The reason for the 
underlying increase of $\tauh$ with $\dot m$ is that the increase of
$\Hd$ with $\dot m$ is slower than linear. The gravity parameter 
$A \propto \Hd/{\dot m}$ then decreases with $\dot m$. We note that the 
vertically averaged SS solution predicts $\Hd\propto {\dot m}$ and,
in consequence, constant $\tauh$ in the regime of $\Prad \gg \Pgas$ and 
$\kappa \approx \kappa_{\rm es}$. In our vertically explicit solution 
$\Pgas$ is not neglected. Contribution from convective energy transfer
contributes to departures from the SS solution as well.

\section{Discussion}
\label{sec:discus}

Presence of a hot, ionized layer on top of an X--ray illuminated accretion 
disc has inevitable observational consequences. Most importantly, it reduces
the amplitude of the effective, observed ``cold'' reprocessed component. Since 
the thickness of the hot layer (hence the reduced amplitude) is correlated
with the X--ray flux, a number of observable correlations follow from
this model.

Generally, the predicted Fe \Ka fluorescent line originating from the disc 
is weaker (relative to the continuum) when the source is brighter
(provided that the illuminating spectrum has $\Gamma<2$ so that
there is no He-like Fe in the hot layer; NKK).
This effect can be used to explain two observed phenomena.
Firstly, the reduction of the amplitude of variability of 
the line with respect to the driving continuum ($E>7$ keV) in AGN monitoring
campaigns (Reynolds 2000; Done et al.\ 2000; Chiang et al.\ 2000). 
Quantitatively,
our computations indicate that it is not possible to obtain constant absolute 
line flux constant. That is, the relative amplitude of the reprocessed
component decreases more slowly than $\FX^{-1}$. Nevertheless, reduction
of the variablity amplitude by a few per cent is possible. However,
the variability reduction is inversly related with the mean amplitude of
the reprocessed component (i.e.\ observed at the mean flux level). Thus,
in sources like MCG--6-30-15, where $R \approx 1$, this effect should actually
be negligible. Moreover, the X--ray continuum in \mcg is rather soft
($\Gamma \ga 2$), which implies a contribution from a highly ionized
reprocessed component (NKK), that does not seem to have been observed in
this source. Perhaps the hot skin developes on accretion discs in the
other Seyfert 1 galaxies, where both $R<1$ and $\sim$constant Fe \Ka line 
flux was observed (IC~4329a, Done et al.\ 2000; NGC~5548, 
Chiang et al.\ 2000). Quality of current \xte data should be good enough
to quantitatively test the model relation presented in 
Fig.~\ref{fig:reductvar}.

Second, the decrease of the EW of the Fe \Ka line in brighter sources
(X--ray ``Baldwin effect'') can be explained, if the inverse relation between
efficiency of the line production and X--ray flux is applied to a sample
of objects. This is not quite a trivial statement, since different average
X--ray luminosities of sources are rather due to different mass accretion 
rates, than a variable output of some magnetic activity.

The model is also able to explain the correlation between the amplitude of
the reprocessed component and X--ray spectral index, again observed
both in a sample of objects (Zdziarski et al.\ 1999), and in the time
history of a single object (GX~339-4, Gilfanov et al.\ 2000; 
GS~1124-68, \.{Z}ycki et al.\ 1998). Our computations of the correlation
give perhaps somewhat different shape to what is observed, although
the observed correlation is also subject to instrument--related
uncertainties.

Another  relation often looked for in data and used to constrain 
theoretical models is a relation between spectral index and hard 
X--ray flux. This is usually more
difficult to obtain from data because of uncertain corrections from
limited detector band pass to the entire X--ray band. Perhaps the most
robust result was obtained by Done et al.\ (2000) for IC~4329a, where
broad band spectral analysis revealed that the spectrum became harder
when the X--rays ($E>1$ keV) decreased. Similar relation was reported for
NGC~5548 (Magdziarz et al.\ 1998; Chiang et al.\ 2000) although this 
involved flux in 2--10 keV band.
On the other hand the soft X--ray transient GS~1124-68 (Nova Muscae 1991) 
showed a clear hardening of its spectrum during its decline phase 
(\.{Z}ycki et al.\ 1998; \.{Z}ycki 2001),
which was however correlated with an {\it increase\/} of the 
$E>1\,{\rm keV}$ flux by $\approx 50$ per cent (so called 're-flare').
The model discussed in this work predicts anti-correlation of 
$\Gamma$ and $\Fhard$, i.e.\ hardening of the spectrum when a source brightens,
unless there are correlated changes of the X--ray source geometry and its
luminosity.

The structure of the hot layer and resulting correlations between
observables may also depend upon the model adopted for the underlying
accretion disc. In this paper we adopted the $\alpha\Ptot$ SS prescription for
viscosity. However such discs are subject to a number of instabilities 
in the parameter range appropriate to AGN. Classical theory predics
thermal and viscous instability whenever contribution of gas
pressure to $\Ptot$ drops below 0.4 (Kato, Fukue \& Mineshige 1998),
which in practice means the region within a few hundreds $\Rg$, for
$\dot m$ a few per cent. Therefore, models with a different
viscosity prescription were sometimes considered, e.g.\ $\alpha\Pgas$,
which are free from those instabilities. Solving the 
vertical structure of $\alpha\Pgas$ discs for a number of ($\xi,\ \eta$)
values  we find that the disc height is very similar (less than 1 per cent
difference) to that of the corresponding $\alpha\Ptot$ disc. The resulting
relations between observable quantities are then practically identical
to those for $\alpha\Ptot$ discs.
Our current understanding of mechanisms providing viscosity
strongly points out towards the magneto-rotational instability
as the source of turbulence responsible for the viscosity 
(see Balbus \& Hawley 1998 for review).
Global magnetohydrodynamical simulations reveal the disc structure
very different from the smooth, classical, hydrodynamical models
(Hawley 2000). The spatial and temporal
structure is highly turbulent and inhomogeneous, displaying fractal-like
scalings (Kawaguchi et al.\ 2000). The relevance of the presented model
to such inhomogeneous, clumpy discs needs further studies.

The presented model is currently one of at least three general scenarios
within which it is possible to explain bulk of observed data from low/hard
states of AGN and GBH: the other two are the cold disc with inner hot
flow model (Esin et al.\ 1997 and references therein; R\'{o}\.{z}a\'{n}ska
\& Czerny 2000b) and relativistically
outflowing plasma model (Beloborodov 1999ab). None of them fully addresses
all complexities of observed spectral and timing data, but all three
account for the main facts: the range of observed spectral indices, 
the $R$--$\Gamma$ correlation, relativistic smearing of Fe reprocessed
features (see Poutanen 1998 for review and e.g.\ Di Salvo et al.\ 2001 
for application to Cyg X-1). 
It also seems possible to incorporate into all of them the basic X--ray timing 
observables: the shape of PSD, time lags, auto- and cross-correlation
functions (Poutanen \& Fabian 1999; B\"{o}ttcher \& Liang 1999; 
see review in Poutanen 2000).
Clearly, further progress can only be made if the models are developed
in more details incorporating more relevant physical processes.
Confronting detailed, quantitative predictions with good quality
spectral and timing data should hopefully enable distinguishing which,
if any, of our current ideas are closest to reality.


\section{Conclusions}

\begin{itemize}
 \item The model does predict certain reduction of the variability
   of the Fe \Ka line, compared to the variability of its driving
   continuum, although it does not seem possible to obtain
   absolutely constant line flux.
 \item The anti-correlation of the equivalent width of the \Ka line
  and average source luminosity (the X--ray ``Baldwin effect'') is 
  reproduced in the model.
 \item Reduction of the amplitude of the observed, ``cold'' reprocessed
   component is more evident than any reduction of the spectral broadening
   (due to relativistic effects and Comptonization) of the \Ka line.
 \item The model is able to quantitatively explain the observed $R$-$\Gamma$
  correlation.
 \item The geometry of the X--ray sources is quite tightly constrained,
 if the model is to explain both timing and spectral observational data.
\end{itemize}

\section*{Acknowledgments} 
 
We acknowledge helpful discussions with Bo\.{z}ena Czerny, Chris Done,
Sergei Nayakshin and Grzegorz Wardzi\'{n}ski.
This work  was supported in part by grants no.\ 2P03D01816 and 2P03D01718  
of the Polish State Committee for Scientific Research (KBN). 

{}

\bsp

\end{document}